# Solvation of Lithium Ion in Dimethoxyethane and Propylene Carbonate

Vitaly Chaban[1]

1) Instituto de Ciência e Tecnologia, Universidade Federal de São Paulo, 12231-280, São José dos Campos, SP, Brazil

2) Department of Chemistry, University of Southern California, Los Angeles, CA 90089, United States

**Abstract**. Solvation of the lithium ion ($Li^+$) in dimethoxyethane (DME) and propylene carbonate (PC) is of scientific significance and urgency in the context of lithium-ion batteries. I report PM7-MD simulations on the composition of $Li^+$ solvation shells (SH) in a few DME/PC mixtures. The equimolar mixture features preferential solvation by PC, in agreement with classical MD studies. However, one DME molecule is always present in the first SH, supplementing the cage formed by five PC molecules. As PC molecules get removed, DME gradually substitutes vacant places. In the PC-poor mixtures, an entire SH is populated by five DME molecules.

[1] E-mail: vvchaban@gmail.com

TOC Graphic

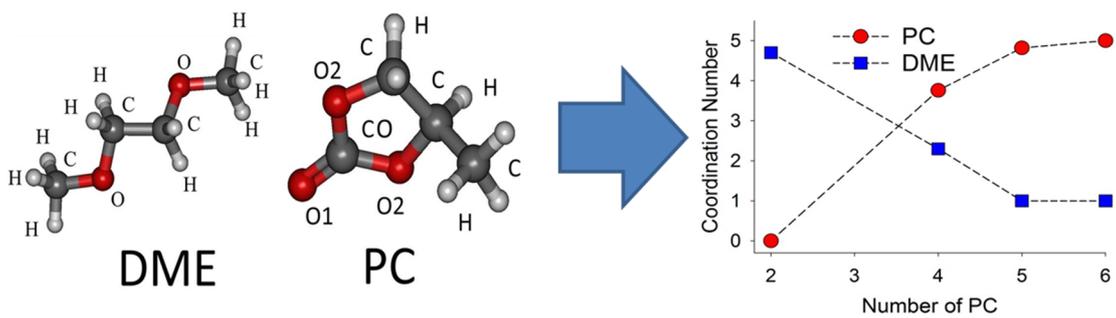

**Research Highlights**

1) PM7-MD simulations have been conducted to study solvation of lithium in propylene carbonate and dimethoxyethane.

2) Binding with PC is significantly more favorable thermodynamically for lithium than binding with DME.

3) The lithium first solvation shell strongly depends on the concentration of the DME/PC mixture.

**Introduction**

Search for novel and investigation of the existing electrolyte solutions[1-8] in organic solvents constitute a great importance and urgency in the today's physical chemistry and electrochemistry due to the recent emergence of multiple portable devices. Lithium salts[9-12] dissolved in the mixed molecular solvents, such as dimethoxyethane (DME) and propylene carbonate (PC) are applied in the lithium-ion batteries,[13-15] since they exhibit wide electrochemical windows and other favorable properties. High ionic conductivity and dielectric constant of the electrolyte solution always arrive at the cost of a high shear viscosity.[14] The compromise can be achieved by mixing highly polar cyclic ethers, such as PC, with low-polar linear ethers, such as DME. Fortunately, these solvents are well miscible with one another giving wise to a variety of compositions. The resulting mixtures provide an electrochemically interesting background for electrolyte systems.

The complexity of experimental setups and molecular-level data interpretations in the case of many-atom molecules considerably hinders development of the efficient electrolyte solutions. The factors determining key properties, such as ion solvation regularities and ion transport regularities, must be clearly understood to foster progress in the lithium-ion batteries. Mixed solvents are challenging for theoretical investigation because of the abundant specific molecular interactions in these systems. The complexity increases drastically with every new molecular or ionic species added to the mixture.

No consensus still exists in the community regarding the structure of the lithium ion solvation shells in its DME-PC solutions. The coordination number of $Li^+$ in the solutions remains actually unknown. Depending on the method of investigation and working approximations, the coordination number is reported anywhere between two and eight.[14,16-20] While the coordination number of eight is unlikely physical due to the size of the solvent molecules, the coordination number of two suggests virtually non-solvated ion in the polar solvent, which is bizarre. Better precision is required to build theoretical descriptions of these ion-molecular systems.

Prezhdo and coworkers[14] have recently reported classical molecular dynamics (MD) simulations to study lithium tetrafluoroborate (LiBF$_4$) in the pure and mixed cyclic and linear carbonates with various salt concentrations. The additive, non-polarizable force field was applied to describe ion-ionic, ion-molecular, and molecule-molecular interactions in these complex systems at room conditions. They found that Li$^+$ is preferentially solvated by a cyclic and more polar co-solvent, suggesting that a strong electrostatic attraction dominates over possible steric hindrances. The coordination number of Li$^+$ varies from 5 to 6 depending on the lithium salt concentration.[14] This variation is observed due to formation of ion pairs in the concentrated solutions of LiBF$_4$. Unfortunately, the model applied by these authors does not account for electronic polarization, which definitely occurs in the real-world systems between Li$^+$ and oxygen atoms of the co-solvents, especially with those of PC. This con of the interaction model may turn out critical at certain circumstances.

The polarizable model by Borodin and Smith[8] provides the first peak for the Li-O atom pair (in LiPF$_6$–ethylene carbonate) near 2.0 Å with a height of 30, which is closer to the result of the ab initio molecular dynamics (AIMD) simulation[7] and significantly smaller than the results obtained using the non-polarizable models.[14] The authors suggest that the existing non-polarizable models overestimate binding energy between the lithium-ion and cyclic carbonates. This supposition looks, in general, reasonable due to fixed electrostatic charges employed in the discussed pairwise models. Ganesh and coworkers[7] reported AIMD simulations of ethylene carbonate and propylene carbonate with LiPF$_6$ at the experimentally relevant concentrations to build solvation models. These developed models can be used to explain available neutron scattering and nuclear magnetic resonance results and to compare the lithium-ion solvation energies and self-diffusion constants. The Born-Oppenheimer AIMD empowered by the pure exchange-correlation functional was used with specific pseudopotentials in the canonical and microcanonical ensembles. According to these simulations, the coordination number of Li$^+$ equals to four,[7] which is smaller than the above introduced result (5-6) by Prezhdo and coworkers.[14]

This persisting controversy in literature calls for an additional attention to the problem of Li$^+$ solvation in the electrolyte systems. On one side, the AIMD simulations offer a more physically relevant and mathematically sophisticated description of the phenomenon. On another side, the pure density functionals are well-known to overestimate electron transfer, which may result in the underestimation Li-solvent binding energy (in contrast to possible overestimation in the study or Prezhdo et al.)[14] Furthermore, small systems used in AIMD cannot properly account for a long-range structure, which is important in any ion-molecular system. Finally, the AIMD simulations reported by Ganesh et al. were conducted at 310-400 K. While the coordination number of 4 (Li-O) may be completely relevant at 400 K, it is not fairly clear whether the performed sampling at 310 K is enough to derive properties. Re-organization of the first coordination sphere can require tens of picoseconds at room conditions.

This work contributes the ongoing discussion by performing PM7-MD simulations of a few [Li–DME–PC]$^+$ non-periodic systems to investigate (1) structure and composition of the lithium-ion first and second solvation shells; (2) dependence of the lithium-ion coordination number on the DME-PC mixture content; (3) energetics of the lithium-ion binding to both co-solvent molecules. The PM7-MD method[21-23] provides a comprehensive tool offering an electronic-structure description of every system. Thanks to intelligently parameterized integrals in PM7,[24-27] PM7-MD performs better than density functional theory powered AIMD simulations in the sense of computational cost. Unlike pure density functional theory, PM7 does not have a reputation of a method that overestimates an electron transfer. Nevertheless, comparison with the experimental data, where available, is desirable.

**Methodology**

The PM7-MD simulations[21-23] were performed using the three systems (Table 1) featuring mixtures of propylene carbonate and dimethoxyethane. All simulations were performed at 300 K with temperature maintained constant using weak coupling to the external thermal bath.[28]

Table 1. Simulated systems, their fundamental properties and selected simulation details. Each system was started with three different configurations (arbitrarily arranged molecules) to ensure that the equilibrated ion-molecular configuration does not depend on the starting configuration. Proper equilibration of all systems was thoroughly controlled by analyzing evolution of many thermodynamic quantities, such as energy components, dipole moments, selected interatomic distances. Note that equilibration of the non-periodic systems occurs significantly faster due to absence of the long-order structure

| # | Composition | # atoms | # explicit electrons | Type of calculation | Equilibration time, ps | Sampling time, ps |
|---|---|---|---|---|---|---|
| 1 | 1 DME | 16 | 38 | EF-OPT | — | — |
| 2 | 1 PC | 13 | 40 | EF-OPT | — | — |
| 3 | 1 DME + 1 PC | 29 | 77 | EF-OPT | — | — |
| 4 | 1 Li$^+$ +1 DME | 17 | 38 | MD | 1.0 | 10 |
| 5 | 1 Li$^+$ +1 PC | 14 | 40 | MD | 2.0 | 10 |
| 6 | 6 DME + 6 PC | 175 | 468 | MD | 15 | 75 |
| 7 | 6 DME + 5 PC | 162 | 428 | MD | 15 | 75 |
| 8 | 6 DME + 4 PC | 149 | 388 | MD | 15 | 75 |
| 9 | 6 DME + 2 PC | 121 | 298 | MD | 15 | 75 |

The PM7-MD method obtains forces acting on every atomic nucleus from the electronic structure computation using the PM7 semiempirical Hamiltonian.[24-27] PM7 is a parameterized Hartree-Fock method, where certain integrals are pre-determined based on the well-known experimental data, such as ionization energies. This solution allows for effective incorporation of the electron-correlation effects, while preserving a quantum-chemical nature of the method. Therefore, PM7 is able to capture any specific chemical interaction. On the contrary, classical pairwise interaction potentials are unable to represent formation/destruction of the covalent bonds, whereas formation/destruction of the hydrogen bonds can be modeled using point electrostatic charges. PM7 is more physically realistic than any existing force field based technique. Note that PM7 includes an empirical correction for the dispersive attraction. Thus, it can be successfully used to model hydrocarbon moieties. The accuracy and robustness of the PM7 parameterization, as applied to thousands of versatile chemical systems, was demonstrated by Stewart elsewhere.[24-27]

The derived forces are coupled with the initial positions of atoms and randomly generated velocities (Maxwell-Boltzmann distribution). Subsequently, Newtonian equations-of-motions can be constructed and numerically integrated employing one of the available algorithms. This work relies on the velocity Verlet integration algorithm. This integrator provides a decent numerical stability, time- reversibility, and preservation of the symplectic form on phase space. Due to rounding errors and other numerical inaccuracies, total energy of the system is not perfectly conserved, as in any other MD simulation method. Temperature may need to be adjusted periodically by rescaling atomic velocity aiming to obtain the required value of kinetic energy with respect to the number of degrees of freedom. This work employs a weak temperature coupling scheme with a relaxation time of 50 fs, whereas the integration time-step equals to 0.5 fs. This integration time-step is set based on the preliminary benchmarks on proper energy conservation (using molecular dynamics simulations in the constant-energy ensemble).

More details of the present PM7-MD implementation are provided elsewhere.[21-23] The method has been successfully applied to address problems of ionic liquids[21,22] and nanoparticles.[23] Local structure of the liquid-matter ion-molecular systems was characterized using a set of pair correlation functions (RDFs). The RDFs were calculated using simple in-home tools along the sampling stage of each PM7-MD trajectory (Table 1). Prior to quantum-chemical analysis, geometries of systems I, II, and III were optimized using the eigenfollowing (EF) algorithm (implemented in MOPAC2012 in the improved form).[27]

**Results and Discussion**

The optimized geometries of the DME and PC molecules are depicted in Figure 1. The above molecules are similar in terms of total number of electrons and number of atomistic composition. However, the structure formulas are different justifying their different physical chemical and thermodynamics properties.

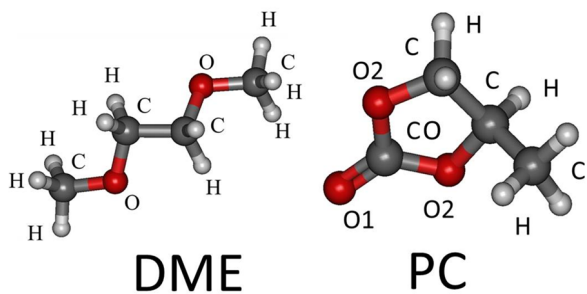

Figure 1. Optimized chemical structures of dimethoxyethane and propylene carbonate molecules. The geometry optimizations were carried out using the EF algorithm and the wave function was constructed using the PM7 Hamiltonian approximation. These designations of atoms will be used throughout the paper to discuss the simulated systems.

As in other ethers, DME molecules are weakly associated and exhibit modest chemical reactivity. DME molecule is large enough to constitute a liquid at room conditions, though. It is a clear, colorless, flammable, volatile, low-viscous substance, which exhibits a good solvation potential. DME is miscible with water. In turn, PC features a carbonate, $CO_3$, group, which is responsible for strong electrostatic interactions and association in the condensed phase. It exhibits a high dielectric constant (64),[14] similarly to other polar aprotic solvents. PC creates an effective solvation shell around alkali ions, thereby creating a conductive electrolyte. Together with DME, PC is used as a solvent in batteries.

According to the Coulson-type population analysis, symmetric oxygen atoms in DME are moderately electron-rich, -0.39e each. Carbon atoms are also slightly negative, -0.06e in the ethane backbone and -0.20e in the methoxy, -O-CH$_3$, moieties. The molecule is neutralized by the ten hydrogen atoms, which are all positively charged, +0.12-0.15e. Interestingly, all three oxygen atoms of PC also carry -0.39e charges, exactly as in the case of DME. The most strongly charged atom in PC is the carbonate carbon, +0.69e. Such an electronic density distribution implies a good miscibility with polar solvents (water, alcohols, organic acids, etc) and a high solvation potential towards inorganic monoatomic ions. The dipole moment of PC, evaluated from the electrostatic potential, equals to 4.6 D, which is in quite a decent agreement with the experimental value, 4.9 D.[14] Note that the experimental dipole moment must be higher due to an electronic polarization in the condensed phase.

Association of DME and PC (Figure 2) in solution is an important characteristic of how this mixture performs for salt solvation. The oxygen atom of DME approaches the hydrogen atom of PC by 2.29 Å. Another oxygen atom of DME simultaneously approaches another hydrogen atom of PC by 2.37 Å. Multiple C-O-H angles in this molecular pair range from 102° to 130°. These interactions can be classified as weak hydrogen bonds. Compare, the typical length of a hydrogen bond in water (that is, involving the same elements) is 1.97 Å. The ideal bond angle depends on the nature (chemical identity and hybridization) of the hydrogen bond donor.

The formation energy of this complex is significant, ca. 914 kJ mol$^{-1}$, suggesting an electrostatic nature of binding. At finite temperature, the observed association energy will obviously somewhat decrease, although the trend will persist. Despite a strong intermolecular binding, the geometry optimization procedure (Figure 2) takes a large number of cycles, ca. 800 cycles, before achieving a stationary value of the formation energy. This complexity probably occurs due to conformational flexibility of DME. At ca. 400$^{th}$ EF cycle, a final formation energy decay of 8 kJ mol$^{-1}$ takes place. This amount of energy is comparable with the one associated with conformational changes.

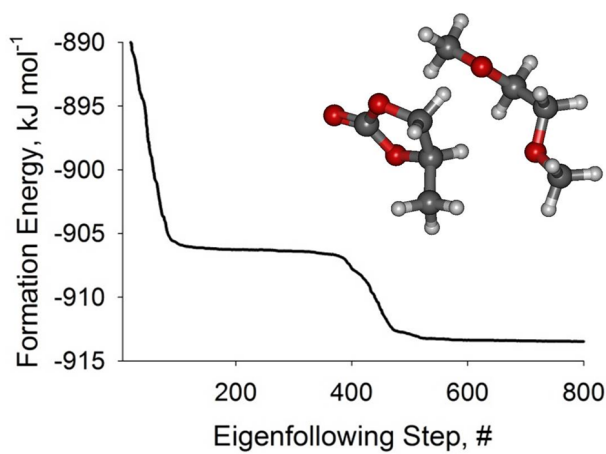

Figure 2. The evolution of formation energy upon geometry optimization of the DME-PC pair of molecules. The geometry optimizations were carried out using the EF algorithm and the wave function was constructed using the PM7 Hamiltonian approximation.

Figure 3 investigates fluctuations of the formation energy and compares the formation energies in the complexes of Li$^+$ with DME and PC separately. The system with PC takes ca. 2 ps to reach thermodynamically equilibrium configuration, whereas the system with DME attains stable energy faster. Fluctuations of the propylene carbonate containing system are less frequent than those corresponding to DME. The performed analysis clearly indicates that the lithium ion prefers to bind with PC. In turn, binding with a single DME molecule appears drastically unfavorable (positive formation energy). Addition of more DME molecules coordinating Li$^+$ stabilizes this complex in vacuum.

Information about an ion behavior in a pure solvent is important to understand physical chemical and thermodynamics trends. However, such information is not sufficient to characterize solvation shells of an ion in the mixed solvent. Solvation is a complicated phenomenon, whereby not only enthalpy (potential energy factor) plays a role, but also entropic contribution to the solvation free energy is extremely important. In the case of sufficiently large molecules, entropic contribution can be even decisive. Based on Figure 3, one can hypothesize that the first solvation shell (FSS) and, probably, the second solvation shell (SSS) of Li$^+$ are populated by the PC molecules, while DME molecules are not located in the vicinity of the cation, due to positive formation energy. In the meantime, we empirically know that DME and PC are miscible well finding routine application in the lithium-ion batteries. In the following, I will illustrate that the solvation shell composition can be only roughly predicted on the basis of enthalpy factor, whereas finite-temperature molecular dynamics simulations are absolutely necessary to observe a realistic distribution of molecules around the Li$^+$ ion.

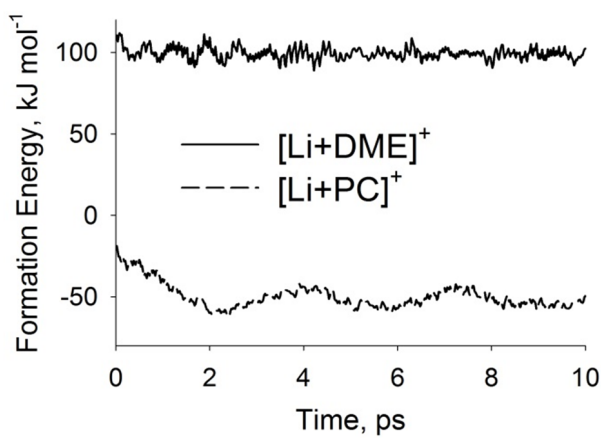

Figure 3. Fluctuations of formation energies in the [Li+DME]$^+$ (solid line) and [Li+PC]$^+$ (dashed line) complexes. See systems IV-V in Table 1. The molecular trajectory was propagated during 10 ps with a time-step of 0.5 fs using the PM7-MD method. Before recording molecular trajectories, the energy of both complexes was minimized using the EF algorithm. The depicted plots allow to conclude that interaction of Li$^+$ with PC is much more favorable than interaction of Li$^+$ with DME in the one-molecule approximation.

Figures 4-5 summarize the most informative pair correlation functions in the [Li–DME–PC]$^+$ systems computed in systems VI-IX. Figure 6 represents the FSS coordination numbers of DME and PC separately. The coordination numbers were derived from the pair correlation functions by taking an integral up to the first minimum. The distinct distribution of the DME molecules (Figure 4) between the first and the second coordination spheres is somewhat unexpected. Both spheres are well-defined, unlike it is observed for PC, where the first minimum is articulated not so clearly (Figure 5). Both DME and PC molecules are present in each coordination sphere. The average separation between Li$^+$ and oxygen atoms constituting the neighboring DME and PC molecules equals to 2.1 Å. This distance is somewhat smaller than the closest-approach distance between PC and DME, as evidenced above. Generally similar correlation functions are observed irrespective of the mixture composition. The only exception is system IX, which does not contain any PC molecule in the FSS (Figure 5). Both of them constitute SSS.

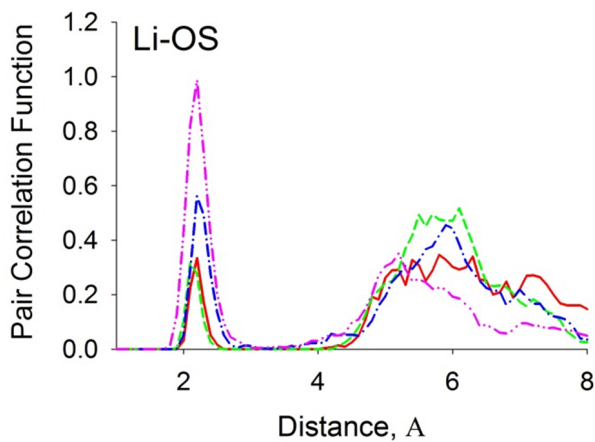

Figure 4. Pair correlation functions derived between lithium ion and oxygen (O) atoms of dimethoxyethane in systems VI, VII, VIII, IX. The functions were computed using the equilibrium parts of the trajectories (Table 1). See legends for designation.

The PC-rich solutions are omitted in this investigation, since preferential solvation of $Li^+$ by the PC molecules is expected based on the data of Figure 3. In addition, DME is added to PC to decrease the shear viscosity (undesirable feature for lithium-ion batteries) and significant contents of the co-solvent are required to achieve that goal. The primary research interest is to unveil how $Li^+$ is solvated in the equimolar and DME-rich mixtures. The FSS of $Li^+$ in the equimolar DME/PC mixture is dominated by the five PC molecules (Figure 6). However, one DME molecule always complements the cage formed by PC. The composition of the $Li^+$ shell is same both in system VI and VII. That is, as long as five PC molecules are available per one lithium ion, the shell is stabilized. The coordination sphere consisting of 5 PC and 1 DME molecules represents the most natural formation in the DME/PC equimolar mixtures. This conclusion confirms results of Prezhdo and coworkers[14] and disagrees with the result of Ganesh and coworkers.[7] Note that unlike classical MD studies, PM7-MD accounts for all specific interactions. I believe that pure density functional of Ganesh et al. unreasonably overestimates an electron transfer and hence underestimates binding energy between Li and O. Even though larger-scale MD simulations may overestimate binding energy and the radial distribution function peak height, the resulting coordination number of lithium appears trustworthy. Upon further removal of PC, DME molecules approach lithium. Acting as bidentate ligands, DME occupies more space near Li$^+$ than PC. Thus, the total number of solvent molecules (PC+DME) decreases when DME enters the FSS (Figure 6).

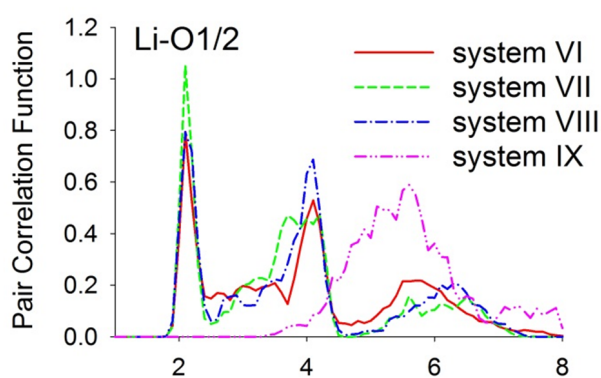

Figure 5. Pair correlation functions derived between lithium ion and oxygen (O1, O2) atoms of propylene carbonate in systems VI, VII, VIII, IX. The functions were computed using the equilibrium parts of the trajectories (Table 1). See legends for designation.

The carbonate group is more electron-rich, which helps it to efficiently coordinate an electron-deficient lithium ion. In turn, DME is a bidentate ligand with the two methoxy groups, but smaller negative electron density on the oxygen atoms. With both molecules being larger than Li$^+$, the FSS composition is guided, in addition, by steric considerations. A single DME molecule constitutes a proof of that, since – otherwise – FSS would have consisted of exclusively PC molecules, as long as enough supply of them is available.

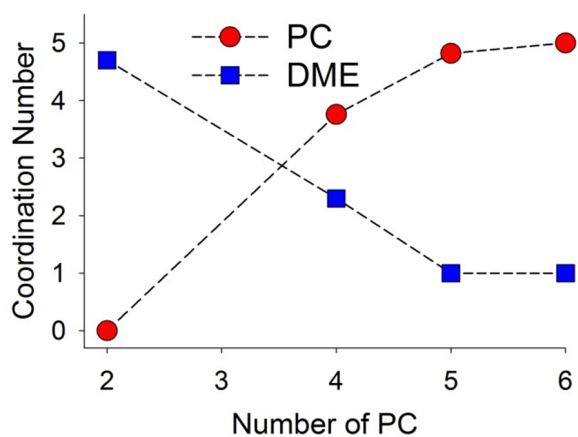

Figure 6. The FSS coordination numbers of the lithium ion as a function of a number of propylene carbonate molecules in the system. The coordination numbers are provided separately for PC (red circles) and DME (blue squares).

**Conclusions**

The paper reports PM7-MD simulations on a set of the [Li–DME–PC]$^+$ systems to understand ion solvation regularities. Ion solvation phenomena are important for an application of these and similar systems as electrolytes in batteries. The PM7-MD method provides an electronic-structure level description including all phenomena, which are inaccessible for classical molecular dynamics. In particular, significant electronic polarization is expected in all systems containing the lithium ion due to its tiny size and unit positive charge. Empirical Hamiltonians either ignore this or represent effectively following an average field approximation. The present PM7-MD results may be perceived as a benchmark for further larger-scale calculations.

The performed simulations confirm that both PC molecules (5) and a single DME molecule contribute the first solvation shell of Li$^+$, although the binding energy of Li$^+$ to a single DME molecule is much weaker than that of PC. This result confirms an accuracy of the recent empirical potential simulations by Prezhdo and coworkers using the simplified additive interaction potentials and the nanosecond time scale trajectories. As PC molecules vanish from the system, DME gradually occupies their places. Interestingly, when the ratio of PC:DME becomes 1:3, propylene carbonate gets completely excluded from the Li$^+$ FSS, despite its favorable binding potential. Five DME molecules substitute all PC molecules, while the latter join SSS. The reported results supports classical MD simulations by Prezhdo and coworkers by confirming their Li$^+$ FSS coordination number[14] and question recent AIMD results, which suggested a smaller coordination number based on shorter trajectories.[7]

The results reported in the paper are important for understanding of the operation of modern lithium-ion batteries. They provide molecular level principles for the optimization of the electrolyte properties, such as solvated ion structure and, further, ionic conductivity and shear viscosity, since the latter depend on the structure and thermodynamics of the solutions.

**Acknowledgments**

The author is CAPES fellow under the "Science Without Borders" program.

**References**

(1)     Tian, W. Y.; Huang, K. M.; Yang, L. J.; Guo, Y. N.; Liu, F. H. Investigate the Microscopic Properties and the Non-Thermal Effect of the Electrolyte Solution under Microwave Irradiation. *Chemical Physics Letters* **2014**, *607*, 15-20.
(2)     Chinchalikar, A. J.; Aswal, V. K.; Kohlbrecher, J.; Wagh, A. G. Evolution of Structure and Interaction During Aggregation of Silica Nanoparticles in Aqueous Electrolyte Solution. *Chemical Physics Letters* **2012**, *542*, 74-80.
(3)     Lima, E. R. A.; Bostrom, M.; Sernelius, B. E.; Horinek, D.; Netz, R. R.; Biscaia, E. C.; Kunz, W.; Tavares, F. W. Forces between Air-Bubbles in Electrolyte Solution. *Chemical Physics Letters* **2008**, *458*, 299-302.
(4)     Ishiyama, T.; Morita, A. Intermolecular Correlation Effect in Sum Frequency Generation Spectroscopy of Electrolyte Aqueous Solution. *Chemical Physics Letters* **2006**, *431*, 78-82.
(5)     Jacob, J.; Kumar, A.; Asokan, S.; Sen, D.; Chitra, R.; Mazumder, S. Evidence of Clustering in an Aqueous Electrolyte Solution: A Small-Angle X-Ray Scattering Study. *Chemical Physics Letters* **1999**, *304*, 180-186.

(6)	Bhatt, M. D.; O'Dwyer, C. Solid Electrolyte Interphases at Li-Ion Battery Graphitic Anodes in Propylene Carbonate (Pc)-Based Electrolytes Containing Fec, Libob, and Lidfob as Additives. *Chemical Physics Letters* **2015**, *618*, 208-213.
(7)	Ganesh, P.; Jiang, D. E.; Kent, P. R. C. Accurate Static and Dynamic Properties of Liquid Electrolytes for Li-Ion Batteries from Ab Initio Molecular Dynamics. *Journal of Physical Chemistry B* **2011**, *115*, 3085-3090.
(8)	Borodin, O.; Smith, G. D. Quantum Chemistry and Molecular Dynamics Simulation Study of Dimethyl Carbonate: Ethylene Carbonate Electrolytes Doped with Lipf6. *Journal of Physical Chemistry B* **2009**, *113*, 1763-1776.
(9)	Barthel, J.; Gores, H. J.; Neueder, R.; Schmid, A. Electrolyte Solutions for Technology - New Aspects and Approaches. *Pure and Applied Chemistry* **1999**, *71*, 1705-1715.
(10)	Kim, H. S.; Kim, H. J.; Cho, W. I.; Cho, B. W.; Ju, J. B. Discharge Characteristics of Chemically Prepared Mno2 and Electrolytic Mno2 in Non-Aqueous Electrolytes. *Journal of Power Sources* **2002**, *112*, 660-664.
(11)	Videa, M.; Xu, W.; Geil, B.; Marzke, R.; Angell, C. A. High Li+ Self-Diffusivity and Transport Number in Novel Electrolyte Solutions. *Journal of the Electrochemical Society* **2001**, *148*, A1352-A1356.
(12)	X, S.; Lee, H. S.; Yang, X. Q.; McBreen, J. A New Additive for Lithium Battery Electrolytes Based on an Alkyl Borate Compound. *Journal of the Electrochemical Society* **2002**, *149*, A355-A359.
(13)	Azeez, F.; Fedkiw, P. S. Conductivity of Libob-Based Electrolyte for Lithium-Ion Batteries. *Journal of Power Sources* **2010**, *195*, 7627-7633.
(14)	Postupna, O. O.; Kolesnik, Y. V.; Kalugin, O. N.; Prezhdo, O. V. Microscopic Structure and Dynamics of Libf4 Solutions in Cyclic and Linear Carbonates. *Journal of Physical Chemistry B* **2011**, *115*, 14563-14571.
(15)	Xu, W.; Shusterman, A. J.; Marzke, R.; Angell, C. A. Limob, an Unsymmetrical Nonaromatic Orthoborate Salt for Nonaqueous Solution Electrochemical Applications. *Journal of the Electrochemical Society* **2004**, *151*, A632-A638.
(16)	Kameda, Y.; Umebayashi, Y.; Takeuchi, M.; Wahab, M. A.; Fukuda, S.; Ishiguro, S. I.; Sasaki, M.; Amo, Y.; Usuki, T. Solvation Structure of Li+ in Concentrated Lipf6-Propylene Carbonate Solutions. *Journal of Physical Chemistry B* **2007**, *111*, 6104-6109.
(17)	Barthel, J.; Buchner, R.; Wismeth, E. Ftir Spectroscopy of Ion Solvation of Liclo4 and Liscn in Acetonitrile, Benzonitrile, and Propylene Carbonate. *Journal of Solution Chemistry* **2000**, *29*, 937-954.
(18)	Henderson, W. A.; Brooks, N. R.; Brennessel, W. W.; Young, V. G. Triglyme-Li+ Cation Solvate Structures: Models for Amorphous Concentrated Liquid and Polymer Electrolytes (I). *Chemistry of Materials* **2003**, *15*, 4679-4684.
(19)	Salomon, M.; Plichta, E. J. Conductivities of 1-1 Electrolytes in Mixed Aprotic-Solvents .2. Dimethoxyethane Mixtures with Propylene Carbonate and 4-Butyrolactone. *Electrochimica Acta* **1985**, *30*, 113-119.
(20)	Soetens, J. C.; Millot, C.; Maigret, B. Molecular Dynamics Simulation of Li(+)Bf4(-) in Ethylene Carbonate, Propylene Carbonate, and Dimethyl Carbonate Solvents. *Journal of Physical Chemistry A* **1998**, *102*, 1055-1061.
(21)	Chaban, V. Competitive Solvation of (Bis)(Trifluoromethanesulfonyl)Imide Anion by Acetonitrile and Water. *Chemical Physics Letters* **2014**, *613*, 90-94.
(22)	Chaban, V. The Thiocyanate Anion Is a Primary Driver of Carbon Dioxide Capture by Ionic Liquids. *Chemical Physics Letters* **2015**, *618*, 89-93.
(23)	Chaban, V. Annealing Relaxation of Ultrasmall Gold Nanostructures. *Chemical Physics Letters* **2015**, *618*, 46-50.
(24)	Stewart, J. J. P. Optimization of Parameters for Semiempirical Methods V: Modification of NDDO Approximations and Application to 70 Elements. *Journal of Molecular Modeling* **2007**, *13*, 1173-1213.
(25)	Stewart, J. J. P. Application of the PM6 Method to Modeling the Solid State. *Journal of Molecular Modeling* **2008**, *14*, 499-535.
(26)	Stewart, J. J. P. Application of the PM6 Method to Modeling Proteins. *Journal of Molecular Modeling* **2009**, *15*, 765-805.

(27) Stewart, J. J. P. Optimization of Parameters for Semiempirical Methods Vi: More Modifications to the NDDO Approximations and Re-Optimization of Parameters. *Journal of Molecular Modeling* **2013**, *19*, 1-32.
(28) Berendsen, H. J. C.; Postma, J. P. M.; Vangunsteren, W. F.; Dinola, A.; Haak, J. R. Molecular-Dynamics with Coupling to an External Bath. *Journal of Chemical Physics* **1984**, *81*, 3684-3690.